\documentclass[aps,prd,twocolumn,showpacs,superscriptaddress,floatfix,nofootinbib,longbibliography]{revtex4}

\usepackage{graphicx}  
\usepackage{dcolumn}   
\usepackage{bm}        
\usepackage{amssymb}   
\usepackage{amsmath}   
\usepackage{epsfig}
\usepackage{subfigure}
\usepackage{psfrag}
\usepackage{multirow}

\usepackage{color}
\usepackage{mathrsfs}

\newcommand{\sigmattbar} {\mbox{\ensuremath{\sigma_{t\bar{t}}}}}
\newcommand{\ttbar}     {\mbox{\ensuremath{t\bar{t}}}}
\newcommand{\ppbar}     {\mbox{\ensuremath{p\bar{p}}}}
\newcommand{\qqbar}     {\mbox{\ensuremath{q\bar{q}}}}
\newcommand{\gq}        {\ensuremath{gq} and \ensuremath{g\bar{q}}}
\newcommand{\dzero}     {D0}
\newcommand{\pt}        {\mbox{$p_T$}}
\newcommand{\met}       {\mbox{\ensuremath{\slash\kern-.55emp_{T}}}}
\newcommand{\metsig}    {\mbox{\ensuremath{\sigma_{\slash\kern-.4emp_{T}}}}}
\newcommand{\ljets}     {\mbox{$\ell$+jets}}
\newcommand{\wjets}     {\mbox{$W$+jets}}
\newcommand{\dilepton}  {\mbox{$\ell\ell$}}
\newcommand{\herwig}    {{\sc herwig}}
\newcommand{\pythia}    {{\sc pythia}}
\newcommand{\alpgen}    {{\sc alpgen}}
\newcommand{\geant}     {{\sc geant3}}
\newcommand{\mcatnlo}   {{\sc mc@nlo}}
\newcommand{\Z}         {\mbox{$Z/\gamma^\star$}}

\newcommand{\etadet}    {\ensuremath{\eta_{\rm det}}}
\newcommand{\msbar}     {\ensuremath{\overline{\mbox{MS}}}}

\newcommand{\lumi}      {9.7~fb$^{-1}$}

\newcommand{\fmeas}     {1.16}
\newcommand{\ferr}      {0.21}

\newcommand{\result} {0.89}
\newcommand{\errstat}{0.16}
\newcommand{\errsys} {0.15}
\newcommand{\errfull}{0.22}

\newcommand{\resultll} {0.80}
\newcommand{\errstatll}{0.22}

\newcommand{\resultlj} {1.02}
\newcommand{\errstatlj}{0.24}

\begin{document}


\noindent%
\mbox{FERMILAB-PUB-15-581-E \hspace{25mm}{\em Published in Phys. Lett. B as 
DOI: 10.1016/J.PhysLetB.2016.03.053}}

\title{Measurement of spin correlation between top and antitop quarks 
produced in  $p\bar{p}$ collisions at $\sqrt{s}=1.96$ TeV}

\affiliation{LAFEX, Centro Brasileiro de Pesquisas F\'{i}sicas, Rio de Janeiro, Brazil}
\affiliation{Universidade do Estado do Rio de Janeiro, Rio de Janeiro, Brazil}
\affiliation{Universidade Federal do ABC, Santo Andr\'e, Brazil}
\affiliation{University of Science and Technology of China, Hefei, People's Republic of China}
\affiliation{Universidad de los Andes, Bogot\'a, Colombia}
\affiliation{Charles University, Faculty of Mathematics and Physics, Center for Particle Physics, Prague, Czech Republic}
\affiliation{Czech Technical University in Prague, Prague, Czech Republic}
\affiliation{Institute of Physics, Academy of Sciences of the Czech Republic, Prague, Czech Republic}
\affiliation{Universidad San Francisco de Quito, Quito, Ecuador}
\affiliation{LPC, Universit\'e Blaise Pascal, CNRS/IN2P3, Clermont, France}
\affiliation{LPSC, Universit\'e Joseph Fourier Grenoble 1, CNRS/IN2P3, Institut National Polytechnique de Grenoble, Grenoble, France}
\affiliation{CPPM, Aix-Marseille Universit\'e, CNRS/IN2P3, Marseille, France}
\affiliation{LAL, Univ. Paris-Sud, CNRS/IN2P3, Universit\'e Paris-Saclay, Orsay, France}
\affiliation{LPNHE, Universit\'es Paris VI and VII, CNRS/IN2P3, Paris, France}
\affiliation{CEA, Irfu, SPP, Saclay, France}
\affiliation{IPHC, Universit\'e de Strasbourg, CNRS/IN2P3, Strasbourg, France}
\affiliation{IPNL, Universit\'e Lyon 1, CNRS/IN2P3, Villeurbanne, France and Universit\'e de Lyon, Lyon, France}
\affiliation{III. Physikalisches Institut A, RWTH Aachen University, Aachen, Germany}
\affiliation{Physikalisches Institut, Universit\"at Freiburg, Freiburg, Germany}
\affiliation{II. Physikalisches Institut, Georg-August-Universit\"at G\"ottingen, G\"ottingen, Germany}
\affiliation{Institut f\"ur Physik, Universit\"at Mainz, Mainz, Germany}
\affiliation{Ludwig-Maximilians-Universit\"at M\"unchen, M\"unchen, Germany}
\affiliation{Panjab University, Chandigarh, India}
\affiliation{Delhi University, Delhi, India}
\affiliation{Tata Institute of Fundamental Research, Mumbai, India}
\affiliation{University College Dublin, Dublin, Ireland}
\affiliation{Korea Detector Laboratory, Korea University, Seoul, Korea}
\affiliation{CINVESTAV, Mexico City, Mexico}
\affiliation{Nikhef, Science Park, Amsterdam, the Netherlands}
\affiliation{Radboud University Nijmegen, Nijmegen, the Netherlands}
\affiliation{Joint Institute for Nuclear Research, Dubna, Russia}
\affiliation{Institute for Theoretical and Experimental Physics, Moscow, Russia}
\affiliation{Moscow State University, Moscow, Russia}
\affiliation{Institute for High Energy Physics, Protvino, Russia}
\affiliation{Petersburg Nuclear Physics Institute, St. Petersburg, Russia}
\affiliation{Instituci\'{o} Catalana de Recerca i Estudis Avan\c{c}ats (ICREA) and Institut de F\'{i}sica d'Altes Energies (IFAE), Barcelona, Spain}
\affiliation{Uppsala University, Uppsala, Sweden}
\affiliation{Taras Shevchenko National University of Kyiv, Kiev, Ukraine}
\affiliation{Lancaster University, Lancaster LA1 4YB, United Kingdom}
\affiliation{Imperial College London, London SW7 2AZ, United Kingdom}
\affiliation{The University of Manchester, Manchester M13 9PL, United Kingdom}
\affiliation{University of Arizona, Tucson, Arizona 85721, USA}
\affiliation{University of California Riverside, Riverside, California 92521, USA}
\affiliation{Florida State University, Tallahassee, Florida 32306, USA}
\affiliation{Fermi National Accelerator Laboratory, Batavia, Illinois 60510, USA}
\affiliation{University of Illinois at Chicago, Chicago, Illinois 60607, USA}
\affiliation{Northern Illinois University, DeKalb, Illinois 60115, USA}
\affiliation{Northwestern University, Evanston, Illinois 60208, USA}
\affiliation{Indiana University, Bloomington, Indiana 47405, USA}
\affiliation{Purdue University Calumet, Hammond, Indiana 46323, USA}
\affiliation{University of Notre Dame, Notre Dame, Indiana 46556, USA}
\affiliation{Iowa State University, Ames, Iowa 50011, USA}
\affiliation{University of Kansas, Lawrence, Kansas 66045, USA}
\affiliation{Louisiana Tech University, Ruston, Louisiana 71272, USA}
\affiliation{Northeastern University, Boston, Massachusetts 02115, USA}
\affiliation{University of Michigan, Ann Arbor, Michigan 48109, USA}
\affiliation{Michigan State University, East Lansing, Michigan 48824, USA}
\affiliation{University of Mississippi, University, Mississippi 38677, USA}
\affiliation{University of Nebraska, Lincoln, Nebraska 68588, USA}
\affiliation{Rutgers University, Piscataway, New Jersey 08855, USA}
\affiliation{Princeton University, Princeton, New Jersey 08544, USA}
\affiliation{State University of New York, Buffalo, New York 14260, USA}
\affiliation{University of Rochester, Rochester, New York 14627, USA}
\affiliation{State University of New York, Stony Brook, New York 11794, USA}
\affiliation{Brookhaven National Laboratory, Upton, New York 11973, USA}
\affiliation{Langston University, Langston, Oklahoma 73050, USA}
\affiliation{University of Oklahoma, Norman, Oklahoma 73019, USA}
\affiliation{Oklahoma State University, Stillwater, Oklahoma 74078, USA}
\affiliation{Oregon State University, Corvallis, Oregon 97331, USA}
\affiliation{Brown University, Providence, Rhode Island 02912, USA}
\affiliation{University of Texas, Arlington, Texas 76019, USA}
\affiliation{Southern Methodist University, Dallas, Texas 75275, USA}
\affiliation{Rice University, Houston, Texas 77005, USA}
\affiliation{University of Virginia, Charlottesville, Virginia 22904, USA}
\affiliation{University of Washington, Seattle, Washington 98195, USA}
\author{V.M.~Abazov} \affiliation{Joint Institute for Nuclear Research, Dubna, Russia}
\author{B.~Abbott} \affiliation{University of Oklahoma, Norman, Oklahoma 73019, USA}
\author{B.S.~Acharya} \affiliation{Tata Institute of Fundamental Research, Mumbai, India}
\author{M.~Adams} \affiliation{University of Illinois at Chicago, Chicago, Illinois 60607, USA}
\author{T.~Adams} \affiliation{Florida State University, Tallahassee, Florida 32306, USA}
\author{J.P.~Agnew} \affiliation{The University of Manchester, Manchester M13 9PL, United Kingdom}
\author{G.D.~Alexeev} \affiliation{Joint Institute for Nuclear Research, Dubna, Russia}
\author{G.~Alkhazov} \affiliation{Petersburg Nuclear Physics Institute, St. Petersburg, Russia}
\author{A.~Alton$^{a}$} \affiliation{University of Michigan, Ann Arbor, Michigan 48109, USA}
\author{A.~Askew} \affiliation{Florida State University, Tallahassee, Florida 32306, USA}
% ...
%
%
\collaboration{The D0 Collaboration\footnote{with visitors from
$^{a}$Augustana College, Sioux Falls, SD, USA,
$^{b}$The University of Liverpool, Liverpool, UK,
$^{c}$DESY, Hamburg, Germany,
$^{d}$CONACyT, Mexico City, Mexico,
$^{e}$SLAC, Menlo Park, CA, USA,
$^{f}$University College London, London, UK,
$^{g}$Centro de Investigacion en Computacion - IPN, Mexico City, Mexico,
$^{h}$Universidade Estadual Paulista, S\~ao Paulo, Brazil,
$^{i}$Karlsruher Institut f\"ur Technologie (KIT) - Steinbuch Centre for Computing (SCC),
D-76128 Karlsruhe, Germany,
$^{j}$Office of Science, U.S. Department of Energy, Washington, D.C. 20585, USA,
$^{k}$American Association for the Advancement of Science, Washington, D.C. 20005, USA,
$^{l}$Kiev Institute for Nuclear Research, Kiev, Ukraine,
$^{m}$University of Maryland, College Park, MD 20742, USA,
$^{n}$European Orgnaization for Nuclear Research (CERN), Geneva, Switzerland
and
$^{o}$Purdue University, West Lafayette, IN 47907, USA.
$^{\ddag}$Deceased.
}} \noaffiliation
\vskip 0.25cm


\begin{abstract}
We present a measurement of the correlation between the spins of $t$ and $\bar{t}$ quarks 
produced in  proton-antiproton collisions  at the Tevatron Collider at  a center-of-mass energy of 1.96~TeV.
We apply a matrix element technique to dilepton and single-lepton+jets final states
in data accumulated with the \dzero\ detector 
that correspond to an integrated luminosity of \lumi.
The measured value of the correlation coefficient in the off-diagonal basis,
$O_{\rm{off}} = \result\pm \errfull\ (\rm{stat} + \rm{syst})$,
is in agreement with the standard model 
prediction, and represents evidence for a top-antitop quark spin correlation
difference from zero at a level of 4.2 standard deviations.
\end{abstract}


\maketitle


\section{Introduction}
\label{Introduction}
The top quark is the heaviest elementary
particle in the standard model (SM)~\cite{Abazov:2014dpa,Abazov:2015spa,Tevatron:2014cka,ATLAS:2014wva}.
Despite the fact that the top quark decays weakly, its large mass leads to a  very short lifetime
of $\approx 5\cdot10^{-25}$~s~\cite{Jezabek:1987nf, Jezabek:1988iv,Abazov2012n}.
It decays to a $W$ boson and a $b$ quark before hadronizing, 
a process that has a characteristic time of
$1/\Lambda_{\rm{QCD}} \approx (200~{\rm MeV})^{-1}$
equivalent to $\tau_{\rm{had}} \approx 3.3\cdot10^{-24}$~s,
where $\Lambda_{\rm{QCD}}$ is the fundamental scale of
quantum chromodynamics (QCD).  The top quark lifetime is also smaller
than the spin-decorrelation time from spin-spin interactions with the
light quarks generated in the fragmentation process~\cite{PhysRevD.49.3320},
$\tau_{\rm{spin}} \approx m_t/\Lambda_{\rm{QCD}}^2 \approx (0.2~{\rm MeV})^{-1} \approx 3\cdot 10^{-21}$~s~\cite{Willenbrock:2002ta}.
The top quark thus provides a unique opportunity
to measure spin-related phenomena  in the quark sector 
by exploiting kinematic properties of its decay products.

In proton-antiproton (\ppbar) collisions,
the dominant process for producing top quarks is through
top-antitop (\,$t\bar{t}$\,) quark pairs.
This QCD process  yields  unpolarized  $t$ and $\bar{t}$ quarks,
but leaves the spins of $t$ and $\bar{t}$ correlated.
A spin correlation observable can be defined as~\cite{Bernreuther:2010ny}
\[
\begin{split}
O_{ab} & = \langle 4(S_t\cdot \hat{a}) (S_{\bar{t}} \cdot \hat{b}) \rangle = \\
 &  \frac{\sigma(\uparrow\uparrow)+\sigma(\downarrow\downarrow)-\sigma(\uparrow\downarrow)-\sigma(\downarrow\uparrow)}
     {\sigma(\uparrow\uparrow)+\sigma(\downarrow\downarrow)+\sigma(\uparrow\downarrow)+\sigma(\downarrow\uparrow)} ,
\end{split}
\]
where $S$ is a spin operator, $\hat{a}, \hat{b}$ are the spin quantization axes for the top
quark ($\hat{a}$) and the antitop quark
($\hat{b}$), $ \langle \rangle$ refers to an expectation value,
$\sigma$ is the \ttbar\ production cross section,
and the arrows refer to the spin states of the $t$ and $\bar{t}$ quarks relative to the 
$\hat{a}$ and $\hat{b}$ axes.
The strength of the correlation depends 
on the \ttbar\ production mechanism~\cite{Mahlon:1995zn,Mahlon:1997uc,Bernreuther:2001rq}.
In  \ppbar\ collisions at a center-of-mass energy of 1.96~TeV, the correlation of spins is predicted 
to be $O_{\rm off} = 0.80^{+0.01}_{-0.02}$~\cite{Bernreuther:2010ny} in the off-diagonal spin basis,
the basis in which the strength of the spin correlation is maximal at the Tevatron~\cite{Mahlon:1997uc}.
The most significant contribution is from the
quark-antiquark annihilation process ($q\bar{q} \to t\bar{t}$) with a spin correlation
strength of $\approx 0.99$, while the gluon-gluon ($gg$) fusion process ($gg \to t\bar{t}$)
has anticorrelated spins with a typical strength of $ \approx -0.36$ at 
next-to-leading order (NLO) in QCD \cite{Bernreuther:2004jv,Bernreuther:2010ny,Bernreuther_private}.
Contributions to \ttbar\ production from beyond the SM  can  have different dynamics that
affect the strength of the  \ttbar\ spin correlation.

Evidence for \ttbar\ spin correlations based on a matrix element technique~\cite{Abazov2012f},
was presented by the \dzero\ collaboration.
 Earlier lower precision measurements used a template method~\cite{Aaltonen:2010nz,Abazov2011aj}.
Spin correlation effects have also been measured in  proton-proton ($pp$) collisions by two
LHC collaborations, ATLAS and CMS,  at a center-of-mass energy of 7~TeV
\cite{ATLAS:2012ao,Aad:2014pwa, Aad:2015bfa, Chatrchyan:2013wua}
and at 8~TeV \cite{Aad:2014mfk, Khachatryan:2015tzo} .
The main mechanism for \ttbar\ production at the LHC is the 
$gg$ fusion process. The spin correlation at the LHC 
arises mainly from the fusion of like-helicity gluons~\cite{Mahlon:2010gw}.
The differences between $pp$ and \ppbar\ incident channels,
the different sources of spin correlation
(quark-antiquark annihilation  versus like-helicity $gg$ fusion),  
and  their different collision  energies, make the measurements of the 
strength of the spin correlation  at both the Tevatron and LHC  interesting and complementary.

In this letter, we present an updated measurement of the \ttbar\ spin correlation strength in
\ppbar\ collisions at $\sqrt s = 1.96$~TeV. 
The measurement uses the statistics accumulated during 2001 -- 2011  data taking period of 
the Fermilab Tevatron Collider, which corresponds to an integrated luminosity of 
\lumi, which is almost two times more than in our previous publication~\cite{Abazov2012f}.


\section{Detector, Event Selection and Simulation, Background}
\label{sec:detector}

The \dzero\ detector is described  in Refs.~\cite{Abazov2006l,Abazov:2005uk,Abolins2008,Angstadt2010,Ahmed:2010fx,Casey:2012rr,Bezzubov:2014jka}. 
It has a  central tracking system consisting of a
silicon microstrip tracker and a central fiber tracker,
both located within an $\sim 2$~T superconducting solenoidal
magnet. The central tracking system is designed to optimize tracking and
vertexing at detector pseudorapidities  of 
$|\etadet|<2.5$\footnote{The pseudorapidity is 
defined as $\eta = - \ln [\tan(\theta/2)]$, where $\theta$ is the polar 
angle of the reconstructed particle originating from a the $p\bar{p}$ collision  vertex,
relative to the proton beam direction.
Detector pseudorapidity $\eta_{\rm det}$ is defined relative to
the center of the detector.}.
The liquid-argon sampling calorimeter has a
central section  covering pseudorapidities $|\etadet|$ up to
$\approx 1.1$, and two end calorimeters  that extend coverage
to $|\etadet|\approx 4.2$, with all three housed in separate
cryostats. A outer muon system, with pseudorapidity coverage of $|\etadet|<2$,
consists of a layer of tracking detectors and scintillation trigger
counters in front of $1.8$~T iron toroids, followed by two similar layers
after the toroids.

Within the SM, the top quark decays with almost 100\% probability into a $W$ boson and a 
$b$ quark.  We also include two final states: the  dilepton final state (\dilepton), where
both $W$ bosons decay to leptons,
and the lepton+jets final state (\ljets), where one of the $W$ bosons
decays into a pair of quarks and one decays to a lepton and a neutrino.
The \ljets\ and \dilepton\ final states contain, respectively,  one or two isolated charged leptons.
In both final states we consider only electrons and muons,
including those from $\tau$-lepton decay, $W \to \tau\nu_\tau \to \ell \nu_\ell \nu_\tau$.
We also require the presence of two $b$ quark jets, two light-quark jets from $W$ decay (in \ljets),
and a significant missing transverse momentum (\met) due to the escaping neutrinos. 

We use the following selection criteria.
In the \dilepton\ channels,  we require two isolated leptons 
with \mbox{$\pt>15$ GeV},
both originating from the same \ppbar\ interaction  vertex.
The \ljets\ channels require one isolated lepton with \mbox{$\pt>20$ GeV}.
We consider electrons and muons identified using the standard \dzero\ criteria 
\cite{Abazov:2013tha,Abazov:2013xpp}, in the pseudorapidity range of 
 $|\etadet|<2.0$ for muons, and $|\etadet|<1.1$ for electrons. In the \dilepton\ channels,
we consider in addition forward electrons in the range of $1.5<|\etadet|<2.5$.
Jets are reconstructed and identified from energy deposition in the calorimeter using
an iterative midpoint cone algorithm~\cite{Blazey:2000qt} of radius~$\sqrt{(\Delta\phi)^2+(\Delta\eta)^2}= 0.5$.
Their energies are corrected using the jet energy scale (JES) algorithm~\cite{Abazov:2013hda}.
All \dilepton\ channels also require the presence of at least two jets with  $\pt>20$~GeV and $|\etadet|<2.5$.
For the  \ljets\ final state, at least four jets must be identified with the same \pt\ and \etadet\
cutoffs, but with the leading jet required to have $\pt>40$~GeV.
When a muon track is found within a jet cone, the JES calculation takes that muon momentum into account, 
assuming that the muon originates from the semileptonic decay of a heavy-flavor hadron belonging to the jet.
To identify $b$ quark jets,
we use a multivariate $b$ quark jet identification discriminant that
combines information from the impact parameters of the tracks and variables that characterize
the presence and properties of secondary vertices within the jet~\cite{Abazov:2013gaa}.
We require that at least one jet is identified as a $b$ quark jet in the \dilepton\ channels, and
at least two such jets in the \ljets\ channels.
To improve signal purity, additional selections based on the global event topology 
are applied~\cite{Abazov:2013wxa,Abazov:2014vga} in each final state. 
A detailed description of event selection can be found in Ref.~\cite{Abazov:2013wxa} for the \dilepton\,  and
in Ref.~\cite{Abazov:2014vga} for the \ljets\ final states.

To simulate \ttbar\ events we use the
next-to-leading (NLO) order Monte Carlo (MC)
QCD generator \mcatnlo~(version~3.4)~\cite{Frixione:2002ik,Frixione:2008ym},
interfaced to \herwig~(version~6.510)~\cite{Corcella:2000bw} for parton
showering and hadronization. The
CTEQ6M parton distribution functions (PDF)~\cite{Pumplin2002,Nadolsky:2008zw} 
are used to generate events at a top quark mass of $m_t=172.5$~GeV.
We use two samples, one including spin correlation effects, and 
the other without correlation.
The generated events are processed through a \geant-based~\cite{geant3}
simulation of the \dzero\ detector.
To simulate effects from additional overlapping \ppbar\ interactions,
``zero bias'' events taken from collider data with an unbiassed trigger
based solely on beam bunch crossings are overlaid on the simulated events.
Simulated events are then processed with the same reconstruction program as data.

In the \dilepton\ channels, the main sources of background are
Drell-Yan production, $q\bar{q}\to\Z\to\ell\ell$, diboson
$WW,\ WZ,\ ZZ$ production, and instrumental background.  The
instrumental background arises mainly from multijet and $(W \to \ell
\nu)$+jets events, in which one jet in \wjets\ or two jets in multijet events are
misidentified as electrons, or where muons or electrons originating
from semileptonic decay of heavy-flavor hadrons appear to be  isolated.
The instrumental background is determined from data, while the other
backgrounds are estimated using MC simulations.  For the
\ljets\ channel, in addition to the Drell-Yan and diboson production,
the contribution from \wjets\ production is estimated from MC
simulation, but normalized to data. Electroweak single top quark production
and \ttbar\ dilepton final states are also considered as background.
The Drell-Yan and $(W \to \ell \nu)$+jets samples are generated with
the leading order (LO) matrix element generator
\alpgen\ (version~v2.11)~\cite{Mangano2003}, interfaced to
\pythia~\cite{Sjostrand2006} (version~6.409, \dzero\ modified
tune~A~\cite{Affolder2002}) for parton showering and hadronization.
Diboson events are generated with \pythia.  More details about
background estimation can be found in
Refs.~\cite{Abazov:2013wxa,Abazov:2014vga}.
Table~\ref{tab:events} shows the number of expected events for each background source and for the signal,
and the number of selected events in
data. The number of the expected \ttbar\ events is normalized to the NLO cross
section of $7.45^{+0.48}_{-0.67}$~pb~\cite{Moch:2008qy}.
The observed number of events in the \ljets\ channel is
higher than the expected, mainly due to an excess in the $\mu$ + jets channel.
The expected and observed number of events are consistent when the systematic uncertainties,
partially correlated between the \ljets\ and \dilepton\ channels, are taken into account.
These uncertainties are of the order of 10\%.
The most important contributions are the integrated luminosity, $b$-quark jet modeling,
uncertainties on the \ttbar\ modeling and uncertainty in the heavy flavor NLO
$K$-factors of the $W$ + jets background in the \ljets\ channel.

\begin{table}
  \begin{center}
    \begin{tabular}[t]{lcccc c|c}
      \hline\hline
      & $Z/\gamma^\star$ & Instrumental  & Diboson &\ttbar&\ Total\ \ &Data  \\
      $e\mu$  & 13.2      & 16.4    & 3.7     &\ 303.4\ \ & 336.7  & 347 \\ 
      $ee$ & 12.2         & 1.8     & 1.9     & 102.4 & 118.3  & 105 \\ 
      $\mu\mu$ & 9.8      & 0.0     & 1.7     & 85.0 & 96.5  & 93 \\ \hline
      
      & $W$+jets & Multijet & Other    &  &   &   \\
      $e$+jets   & 22.7 & 23.1 & 15.3  & 427.4  & 488.6  & 534 \\
      $\mu$+jets & 24.1 & 3.5  & 11.6  & 341.4  & 380.6  & 440 \\ 
      \hline\hline
    \end{tabular}
    \caption{Numbers of expected events, and numbers of events found in data.
      \label{tab:events}}
  \end{center}
\end{table}


\section{Measurement Technique and Results}

Our measurement uses  the same matrix element (ME) approach 
as Refs.~\cite{Abazov2011b,Abazov2012f},  adapted to the spin correlation measurement.
This method consists of calculating the spin correlation discriminant~\cite{Melnikov2011a}
\begin{equation}
\label{eq:R}
  R (x)=\frac{P_{t\bar{t}}(x, {\rm SM})}{P_{t\bar{t}}(x, {\rm SM})+P_{t\bar{t}}(x, {\rm null})}\ , 
\end{equation}
where $P_{t\bar{t}}(x, \mathscr{H})$ is a per-event probability for hypothesis $\mathscr{H}$ for
the vector of the reconstructed object parameters $x$. 
Hypothesis $\mathscr{H}={\rm SM}$ assumes the \ttbar\ spin correlation strength
predicted by the SM,
and $\mathscr{H}={\rm null}$ assumes  uncorrelated spins.
These probabilities are calculated from the integral
\begin{eqnarray}
  P_{t\bar{t}}(x,\mathscr{H}) = \frac{1}{\sigma_{\rm obs}} \int f_{{\rm PDF}}(q_1) f_{{\rm PDF}}(q_2) \times 
  \nonumber \\
\frac{(2\pi)^4 |\mathscr{M}(y,\mathscr{H})|^2}{q_1 q_2 s}W(x,y) d\Phi^6 dq_1 dq_2 .
\end{eqnarray}
Here, $q_1$ and $q_2$ represent the respective fractions of proton and antiproton momentum carried by the initial state partons,
$f_{{\rm PDF}}$ represents the  parton distribution functions, 
$s$ is the square of the \ppbar\ center-of-mass energy,
and $y$ refers to partonic final state four-momenta of the particles.
The detector transfer functions, $W(x,y)$, correspond to the probability to reconstruct
four-momenta $y$ as $x$,
$d\Phi^6$ represents the six-body  phase space, and 
$\sigma_{{\rm obs}}$ is the observed \ttbar\ production cross section, 
calculated  using $\mathscr{M}(\mathscr{H}={\rm null})$,
taking into account the efficiency of the selection.
The same $\sigma_{{\rm obs}}$ is used for $\mathscr{H}={\rm null}$ and $\mathscr{H}={\rm SM}$ hypotheses,
because the difference in observed  cross-sections is small, at the order of percent,
and affects only the separation power of the discriminant $R$.
This calculation uses the LO matrix element 
$\mathscr{M}(y,\mathscr{H})$ for the processes 
$q\bar{q} \to t\bar{t} \to W^+W^- b\bar{b} \to \ell^{\pm}\nu_\ell qq^\prime b\bar{b}$ or
$\ell^+\ell^-\nu_\ell \bar{\nu_\ell} b\bar{b}$,
calculated according to the spin correlation hypothesis $\mathscr{H}$.
The matrix element $\mathscr{M}$ is averaged over the colors and spins of the initial partons,
and summed over the final colors and spins.
For the hypothesis $\mathscr{H}={\rm null}$,
we set the spin correlation part to zero~\cite{Mahlon:1995zn,Mahlon:1997uc}.
In the calculation, we assume perfect measurements of the lepton and jet directions, and 
perfect measurement of electron energy, which reduces the number of dimensions that require integration.
The probability is obtained by integrating over the remaining kinematic variables.
In the \dilepton\ final state, we use the 
top and antitop quark masses,  $W^+$ and $W^-$ boson masses, \pt\ of two jets,
$1/p_T$ for any muons and \pt\ and $\phi$ of the \ttbar\ system as integration variables.
In the   \ljets\ final state, the variables are
the top and antitop quark masses, the mass of the $W$ boson decaying to $q\bar{q}^\prime$, \pt\ of the $d$-type quark jet,
$p_z$ of the leptonically decaying top quark and $1/p_T$ of a muon.
Given the inability to know the flavor of the two quarks from the $W$ boson decay,
or which b-tagged jet originates from the decay of  the top or anti-top quark,
all possible jet-parton assignments are considered and  $P_{t\bar{t}}$ is calculated as the
sum over all the probabilities.

The distributions in the discriminant $R$ of Eq.~(\ref{eq:R})
are calculated for  simulated \ttbar\ events with SM spin correlation and
with uncorrelated spins.  These  and the expected contributions from
the background events are used as templates to fit 
the $R$ distribution in data through a binned maximum-likelihood fit with two free parameters:
the \ttbar\ production cross section $\sigmattbar$, and the measured fraction of
events with the SM spin correlation strength, $f$.

\begin{figure}
\includegraphics[width=0.5\textwidth]{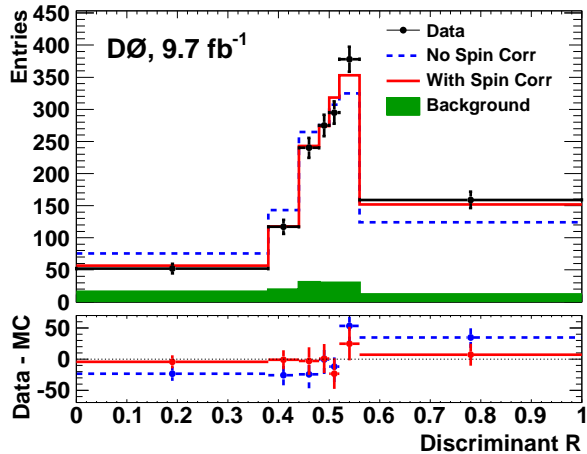}
\caption{Distribution of the spin correlation discriminant $R$ in data and for the \mcatnlo\ \ttbar\ prediction with background,
showing the merged results from  \dilepton\ and \ljets\ events.
The lower plot represents  the difference between data and simulation with SM spin correlation
and without spin correlation.
The error bars correspond to statistical uncertainties.
\label{fig:result}}
\end{figure}

This fit of the distributions in the \dilepton\ and \ljets\ channels is performed simultaneously,
with the expected number of events $n_i$ in each bin $i$ given by
\begin{equation}
  \label{eq:nevt}
  n_i = \frac{\sigmattbar}{7.45 {\rm pb}} \left(f  n_{\rm SM}^i + (1-f) n_{\rm null}^i\right) + n_{\rm bckg}^i,
\end{equation}
where $n_{\rm SM}^i$ and $n_{\rm null}^i$ are the number of events in bin $i$
based on the \mcatnlo\ prediction, with and without spin correlations, and
$n_{\rm bckg}^i$ is the expected number of background events in the same bin.
We use a non-uniform bin width and require a sufficiently large
number of events for each bin in order to avoid bins with zero events, as they
could bias the fit result.
The exact number of bins and their size were optimized to give the smallest
expected statistical uncertainty in the case of the SM spin correlation.
We use the same number and widths of the bins  for the \ljets\
and \dilepton\ channels so as to keep the bin optimization procedure relatively simple.
The fit yields $f = \fmeas \pm \ferr\ (\rm{stat})$.
The $R$ distribution for the combined   \dilepton\ and \ljets\  channels  is shown in Fig.~\ref{fig:result}.
We estimate the significance of the non-zero spin correlation hypothesis 
using the  Feldman and Cousins frequentist
procedure~\cite{Feldman:1997qc}, assuming that the parameter $f$ is in the range  $[0, 1]$, even though the measured
value obtained in the fit is outside of the range $[0, 1]$.

To translate the $f$ value to the 
spin correlation strength in the off-diagonal basis $O_{\rm{off}}$,
we must consider the value of the  spin correlation  strength in the simulation $O^{\rm{MC}}_{\rm off}$.
We choose to obtain this value in the simulated \dilepton\ samples from
the expected value of 
$k_1k_2O^{\rm{MC}}_{\rm{off}} =  -9 \langle\cos\theta_1 \cdot\cos \theta_2\rangle$~\cite{Bernreuther:2004jv}, 
where $\theta_1$ and $\theta_2$ represent angles between the respective direction of a positively and negatively
charged lepton and the spin quantization axes in the $t$ and $\bar{t}$ rest frame.
The parameters $k_1$ and  $k_2$ are the spin analyzing-power coefficients of the top quark
(equal to 1 for leptons at LO in QCD)~\cite{Brandenburg:2002xr}. 
With \mcatnlo, the value calculated for the parton-level 
distributions before any selections is found to equal
$O^{\rm{\mcatnlo}}_{\rm{off}}=0.766$ in the off-diagonal basis. 
The measured spin correlation strength for \ljets\ and \dilepton\ channels is therefore
\[O^{\rm meas}_{\rm{off}} = O^{\rm{\mcatnlo}}_{\rm{off}} \cdot f = \result \pm \errstat\ (\rm{stat})\ , \]
in agreement with the NLO QCD calculation $O_{\rm off} = 0.80^{+0.01}_{-0.02}$~\cite{Bernreuther:2010ny}.
For events in the \ljets\ channel, the result is
\[O^{\rm{\ell+jets}}_{\rm{off}} = \resultlj\pm\errstatlj\ (\rm{stat})\ ,\]
and for \dilepton\ channel the result is 
\[O^{\ell\ell}_{\rm{off}} = \resultll\pm\errstatll\ (\rm{stat})\ .\]

We can reinterpret the measured fraction $f$ as the related measurement of the
spin correlation observable
$O_{\rm{spin}}  = \langle \frac{4}{3}(S_t\cdot S_{\bar{t}} ) \rangle$~\cite{Bernreuther:2010ny}.
This observable characterizes the distribution in the opening angle,
$\varphi$, between the directions of the two leptons in dilepton events or
between the lepton and the up-type quark from the $W$ decay in \ljets\  events,
where the directions are
defined in the $t$ and $\bar{t}$ rest frame:
\begin{equation}
  \frac{1}{\sigma}\frac{d\sigma}{d\cos\varphi}=\frac{1}{2}(1-k_1k_2 O_{\rm{spin}}\cos\varphi).
  \end{equation}
The prediction from the \mcatnlo\ simulation is given by the expectation
value $k_1k_2 O^{\rm{\mcatnlo}}_{\rm{spin}} = -3 \langle\cos \varphi\rangle$  at the parton level,
without any selections,
and found to be $O^{\rm{\mcatnlo}}_{\rm{spin}} = 0.20$.
The value  measured from data is therefore 
\[O^{\rm meas}_{\rm{spin}} = O^{\rm{\mcatnlo}}_{\rm{spin}} \cdot f = 0.23 \pm 0.04 (\rm{stat}) ,\]
consistent with  the NLO QCD calculation of $O_{\rm{spin}} = 0.218\pm0.002$~\cite{Bernreuther:2010ny}.


\section{Systematic Uncertainties}

The estimated  systematic uncertainties are summarized in  Table~\ref{tab:syst}.
These are obtained by replacing the nominal \ttbar\ and background results with modified templates, 
refitting the data and determining the new fraction $f_{\Delta}$. 

We consider several sources of uncertainties in the modeling of the signal. These include
initial-state and final-state radiation, the simulation of hadronization and underlying events, 
the effects of higher-order corrections,
color-reconnection and uncertainty on the top quark mass.
The details of the corresponding samples and parameters 
are discussed in Refs.~\cite{Abazov:2014dpa,Abazov:2015spa}.

For the PDF uncertainty,  we change  the 20 CTEQ6 eigenvectors independently and
add the resulting uncertainties in quadrature.
In modeling both the estimated signal and PDF uncertainties, the event samples have  different fractional
contributions from  $gg$ fusion and \qqbar\ annihilation, and therefore different
spin-correlation strengths.
We take this into account by normalizing the measured fraction to the
spin-correlation strength of the sample $O^{\rm{MC}}_{\rm{off}}$,
in a way similar to that used for the
nominal measurement $O^{\Delta}_{\rm{off}} = f_{\rm{\Delta}} \cdot O^{\rm{MC}}_{\rm{off}}$.

The statistical uncertainty in MC templates is estimated using the ensemble testing technique.
The new ensembles are created through a random generation of a new number of events
in each bin of the MC template assuming a Gaussian distribution in the number of events in the bin.
The same distribution in data is fitted with the modified templates and the dispersion
in the fit results over 1000 ensembles is used as an estimation of the statistical uncertainty in the MC templates.

The uncertainty on identification and reconstruction effects includes
uncertainties on lepton, jet and $b$ tagging identification efficiencies, jet energy resolution and scale corrections, trigger efficiencies, and the luminosity.
The uncertainty in the background contributions includes all uncertainties
that affect the signal-to-background ratio that are not contained in the
previous categories. These uncertainties include uncertainties
in theoretical cross sections for backgrounds,
uncertainty in $Z$ boson $p_T$ distribution,
and uncertainties in instrumental background contributions.

The total absolute systematic uncertainty on the spin correlation observable $O^{meas}_{\rm{off}}$,
calculated as a quadratic sum over all individual sources,  is  $0.15$, as shown in Table~\ref{tab:syst}.
\begin{table}[!htb]
\begin{center}
\begin{tabular}[t]{l c}
\hline\hline
Source & Uncertainty in $O^{meas}_{\rm off}$ \\
\hline
Modeling of signal                  & $\pm 0.135$ \\ 
PDF                                 & $\pm 0.027$ \\ 
Statistical fluctuations in MC      & $\pm 0.026 $ \\ 
Identification and reconstruction   & $\pm 0.032 $ \\ 
Background contribution             & $\pm 0.019 $ \\ 
\hline 
Total                               & $\pm \errsys$ \\ 
\hline\hline
\end{tabular}
\caption{Systematic uncertainties (absolute values) on the spin correlation strength $O^{meas}_{\rm off}$.
\label{tab:syst}}
\end{center}
\end{table}

\section{Spin correlation and  the~$t\bar{t}$~production mechanism}
The strength of the \ttbar\ spin correlation in the SM is strongly dependent on the \ttbar\ production mechanism.
The spin correlation measurement thus provides a way of measuring
the fraction of events produced via $gg$ fusion, $f_{gg}$~\cite{Bernreuther:2001rq}.
The $f_{gg}$ fraction is not well defined
at orders higher than LO QCD.
The difficulty arises from the fact that the cross sections for the
$gq\to t\bar{t}q$ and $g\bar{q}\to t\bar{t}\bar{q}$ processes 
at LO, as well as $gg$ and \qqbar\ production at NLO,
contain a singularity when the final state quark is collinear with the quark in the initial state.
This makes the integration over the phase space divergent~\cite{Nason:1987xz,Bernreuther_private, Mangano_private}. 
In practice, this singularity is absorbed into the definition of the PDF, but the final results 
depend on the scheme used for regularization. 
For the NLO PDF, the \msbar\  scheme is usually preferred. 
The \gq\  contribution at NLO is   
of the order of a few percent~\cite{Bernreuther_private, Bernreuther:2010ny,Bernreuther:2004jv},
and considering that the overall spin correlation strength is $\approx 80\% $,
we neglect these smaller contributions, and determine  $f_{gg}$ from the relation
\[ O = (1-f_{gg}) O_{q\bar{q}} + f_{gg}  O_{gg}\ .\]
Assuming $O_{q\bar{q}} \approx 1$, the gluon fraction becomes
\[f_{gg} \approx  \frac{1-O_{\ \ }} {1-O_{gg}}, \]
where $O$ is the measured value of the total spin correlation strength,
and $O_{gg}$ is the SM value of the spin correlation strength for $gg$ events.

The NLO calculation in the off-diagonal basis using the
CT10 PDF yields $O_{gg}= -0.36\pm 0.02 $~\cite{Bernreuther_private, Bernreuther:2010ny,Bernreuther:2004jv}.
The systematic uncertainty on the observable $O$ can be translated to the uncertainty on the gluon fraction
that includes an additional contribution from the theoretical uncertainty on $O_{gg}$. 
In the  absence of non-SM contributions, the fraction of \ttbar\ events produced through  gluon fusion becomes
\[
  f_{gg} = 0.08 \pm 0.12 (\rm{stat}) \pm 0.11 (\rm{syst})  =  0.08 \pm 0.16 (\rm{stat} + \rm{syst})\ ,
\]
 in agreement with the NLO prediction 
of $ f_{gg} = 0.135$~\cite{Bernreuther_private, Bernreuther:2010ny,Bernreuther:2004jv}.


\section{Summary}

We have presented an updated measurement of \ttbar\ spin correlations with the \dzero\ detector
for an integrated luminosity of \lumi.
The result of the measurement of the strength of the \ttbar\ spin correlation in the off-diagonal basis  is 
\begin{eqnarray*}
O_{\rm{off}} & = & \result\pm \errstat\ (\rm{stat}) \pm \errsys\ (\rm{syst}) \\
& = & \result\pm \errfull\ (\rm{stat} + \rm{syst}).
\end{eqnarray*}
This result is in agreement with the NLO QCD calculation $O_{\rm off} = 0.80^{+0.01}_{-0.02}$~\cite{Bernreuther:2010ny}
and supersedes that reported in Ref.~\cite{Abazov2012f}.
Using the Feldman and Cousins approach for  interval setting~\cite{Feldman:1997qc},
and assuming uncorrelated \ttbar\ spins, we estimate a probability (p-value)
of $2.5\times 10^{-5}$ for obtaining a spin correlation larger than the observed value.
This corresponds to evidence for spin correlation in \ttbar\ 
events at a significance of $4.2$~standard deviations.

In the  absence of non-SM contributions, we use the spin correlation strength measurement to constrain 
the fraction of events produced through  gluon fusion at NLO  QCD and obtain 
\begin{eqnarray*}
  f_{gg} =  0.08 \pm 0.16 (\rm{stat} + \rm{syst})\ .
\end{eqnarray*}
 in good agreement with SM prediction.


\section*{Acknowledgment}
We thank the staffs at Fermilab and collaborating institutions,
and acknowledge support from the
Department of Energy and National Science Foundation (United States of America);
Alternative Energies and Atomic Energy Commission and
National Center for Scientific Research/National Institute of Nuclear and Particle Physics  (France);
Ministry of Education and Science of the Russian Federation, 
National Research Center ``Kurchatov Institute" of the Russian Federation, and 
Russian Foundation for Basic Research  (Russia);
National Council for the Development of Science and Technology and
Carlos Chagas Filho Foundation for the Support of Research in the State of Rio de Janeiro (Brazil);
Department of Atomic Energy and Department of Science and Technology (India);
Administrative Department of Science, Technology and Innovation (Colombia);
National Council of Science and Technology (Mexico);
National Research Foundation of Korea (Korea);
Foundation for Fundamental Research on Matter (The Netherlands);
Science and Technology Facilities Council and The Royal Society (United Kingdom);
Ministry of Education, Youth and Sports (Czech Republic);
Bundesministerium f\"{u}r Bildung und Forschung (Federal Ministry of Education and Research) and 
Deutsche Forschungsgemeinschaft (German Research Foundation) (Germany);
Science Foundation Ireland (Ireland);
Swedish Research Council (Sweden);
China Academy of Sciences and National Natural Science Foundation of China (China);
and
Ministry of Education and Science of Ukraine (Ukraine).

\bibliographystyle{apsrev_custom2}
\bibliography{References}


\end{document}